# Asymmetric discrete random walk and drift-diffusion with unequal jump times, lengths, and probabilities

**Guoxing Lin [1]*, Shaokun Zheng [2]**

1. Carlson School of Chemistry and Biochemistry, Clark University, Worcester, MA 01610

2. Department of Radiology, UMASS Medical School, Worcester, Massachusetts, USA

**ABSTRACT**

Random walk has wide applications in many fields, such as machine learning, biology, physics, and chemistry. Random walk can be discrete or continuous in time and space. Asymmetric random walk could be described by drift-diffusion equation. The discrete asymmetric random walk provides a basis for understanding biased drift-diffusion. However, the current reported theoretical results for discrete random walks do not give a general theoretical treatment for the asymmetry due to unequal jump times. In this paper, the theoretical expressions for asymmetric random walks with unequal jump probabilities, times, and lengths are derived. The obtained theoretical results can be reduced to reported results when jump times are equal. Additionally, discrete random walk simulations are performed to verify the obtained theoretical results. There are good agreements between the theoretical predictions and simulation results.

**Keywords:** asymmetric random walk, drift-diffusion, simulation

*Email:glin@clarku.edu

## I. INTRODUCTION

Random walk and diffusion have broad applications in various fields such as biology, economics, machine learning, physics, and chemistry [1,2,3, 4]. The symmetric random walk is closely related to diffusion equations. In contrast, an asymmetric random walk could be described by drift-diffusion equation [5]. Diffusion constant in materials could be detected by various techniques such as pulsed field gradient (PFG) diffusion nuclear magnetic resonance (NMR) [6,7,8,9], where the spin self-diffusion affects the refocusing of NMR magnetization and thus leads to a decay signal intensity. The diffusion constant plays an important role in many real applications. For instance, the diffusion constant is often employed as a contrast parameter in medical magnetic resonance imaging (MRI) [10]. In NMR study, the NMR exchange time constant and the relaxation rate could be obtained from the phase diffusion coefficients of the spin phase random walk [11, 12]. Additionally, the diffusion coefficient is closely related to the characteristic time constant, a dynamic parameter in the investigated system [13]. Furthermore, the time-dependent diffusion constant reveals the morphology information in the material [13].

Various theoretical results have been reported for asymmetric random walk [5,14,15]. For an asymmetric discrete random walk, Ref. [5] derived the drift-equation for the asymmetric discrete random walk with unequal moving probabilities and jump lengths: $p_l \neq p_r$ $p_l + p_r = 1$, $M_r \geq 0$ where $p_l$ and $p_r$ are the moving probabilities in the two opposite walk directions, such as left and right, and $M_r$ is the ratio of the right jump length to the left jump length; while in Ref. [15], the diffusion coefficient for $p_l \neq p_r$, $p_l + p_r \leq 1$, $M_r = 1$ are obtained. The asymmetric random walk could arise from unequal moving probabilities [5,15] or jump length [5]. However, in real applications, the moving time of left and right jumps could be unequal. For instance, the tracer's asymmetric uphill and downhill transition times are investigated in [16]. Additionally, the spin phase random walk among different NMR chemical exchange sites can be an asymmetric random walk with different jump times, lengths, and probabilities [11], which is difficult to be explained by currently available theoretical results. A general theoretical expression is needed to understand the discrete random walk with unequal jump times, lengths, and probabilities, and it could help us better understand more complicated diffusion processes.

In this paper, the asymmetric random walk with $p_l + p_r \leq 1$, $M_r \geq 0$, $M_t \geq 0$, and $M_e \geq 0$ will be investigated, where $M_t$ is the ratio of the right jump time to the left jump time, and $M_e$ is the ratio of the time staying in an immobile site to the left jump time. The left jump time can be denoted as $\tau$. When $p_l + p_r < 1$, the particle in the random walk is assumed to stay in an immobile site for a duration $\tau_{im}$, and $M_e = \tau_{im}/\tau$. $M_e$ may not equal 1. The discrete random walk simulation is carried out to verify the obtained theoretical expressions. The results could help us understand asymmetric random walks and improve its applications.

## II. THEORY

For an asymmetric random walk, $p_l$ and $p_r$ could be different, and $p_l + p_r \leq 1$ [15]. Meanwhile, the random walk could have different left and right jump lengths, $\epsilon$ and $M_r\epsilon$. Additionally, its left jump time $\tau$, right jump time $M_t\tau$, and the waiting time $M_e\tau$ on the immobile sites could be different. The case that $M_t \neq 1$ can be found in asymmetric phase diffusion describing NMR chemical exchange [11]. Note, $M_t \geq 1$ is required in the theoretical derivation in section **II.B**, which, however, does not lose generality because the case $0 < M_t < 1$ can be treated by switching the "left" and "right" jumps.

Because the derivation of the case with $M_t \neq 1$, and $M_e \neq 1$ is more complicated than that with $M_t = 1$ and $M_e = 1$, we first derive the diffusion coefficient and drift-diffusion equation for the asymmetric random walk with $p_l + p_r \leq 1$, $M_r \geq 0$, $M_t = 1$, and $M_e = 1$ in section **II.A**. Then, the asymmetric random walk with $p_l + p_r \leq 1$, $M_r \geq 0$, $M_t \geq 1$, and $M_e \geq 0$ will be derived in section **II.B**.

### A. $\boldsymbol{p_l + p_r \leq 1, M_r \geq 1, M_t = 1}$, and $\boldsymbol{M_e = 1}$ (equal time)

For $M_t = 1$ and $M_e = 1$, each step of the random walk is independent, and it takes the same time $\tau$ regardless of whether the particle moves in the step. The diffusion coefficient can be conveniently derived based on the mean and variance from the random walk.

The expectation value of the mean displacement for one jump can be obtained as

$$E(\Delta x_i) = p_r M_r \epsilon - p_l \epsilon, \tag{1a}$$

and the expectation value of mean displacement for $n$ jumps can be given by

$$E(X(n)) = E\left(\sum_{i=1}^{n} \Delta x_i\right) = \sum_{i=1}^{n} E(\Delta x_i) = \sum_{i=1}^{n}(p_r M_r \epsilon - p_l \epsilon) = n(p_r M_r - p_l)\epsilon. \tag{1b}$$

The variance can be evaluated as

$$\begin{aligned}
\sigma^2 = var(\Delta X_i) &= p_r\, M_r{}^2 \epsilon^2 + p_l\, \epsilon^2 - (p_r M_r - p_l)^2 \epsilon^2 \\
&= \left[p_r\, M_r{}^2 + p_l - p_r{}^2\, M_r{}^2 + 2p_l p_r M_r - p_l{}^2\right]\epsilon^2 \\
&= \left[p_r(1 - p_r) M_r{}^2 + p_l(1 - p_l) + 2p_l p_r M_r\right]\epsilon^2 \\
&= \left[p_r(1 - p_r - p_l + p_l) M_r{}^2 + p_l(1 - p_l - p_r + p_r) + 2p_l p_r M_r\right]\epsilon^2 \\
&= \left\{p_l p_r (M_r + 1)^2 + p_r[1 - (p_l + p_r)] M_r{}^2 + p_l[1 - (p_l + p_r)]\right\}\epsilon^2,
\end{aligned} \tag{2a}$$

and the mean square of the diffusion distance can be calculated as

$$\langle X(n)^2 \rangle = var(X(n)) = var\left(\sum_{i=1}^{n} \Delta X_i\right) = \sum_{i=1}^{n} var(\Delta X_i) = n\sigma^2. \tag{2b}$$

Based on Eq. (2a), the diffusion constant can be given by [1,8]

$$D = \frac{\sigma^2}{2\tau} = \frac{p_l p_r (M_r + 1)^2 + p_r[1 - (p_l + p_r)] M_r{}^2 + p_l[1 - (p_l + p_r)]}{2\tau}\epsilon^2. \tag{3}$$

When $p_l + p_r = 1$, Eq. (3) reduces to

$$D = \frac{p_l p_r (M_r + 1)^2}{2\tau}\epsilon^2. \tag{4}$$

which reproduces the result in Ref. [5]. When $M_r = 1$, Eq. (3) reduces to

$$D = \frac{4p_l p_r + p_r[1-(p_l+p_r)] + p_l[1-(p_l+p_r)]}{2\tau}\epsilon^2, \tag{5}$$

which reproduces the result in Ref. [15]. When $p_l = p_r$ , Eq. (5) reduces to $D = \frac{(p_l+p_r)\epsilon^2}{2\tau}$, which reproduces results in Ref. [17]; however, it works only when $p_l = p_r$, $M_r$ =1.

The above random walk with $p_l + p_r \le 1$, $M_r \ge 1$, satisfies the drift-diffusion equation:

$$\frac{\partial}{\partial t}\rho(x,t) = D\frac{\partial^2}{\partial x^2}\rho(x,t) - v\frac{\partial}{\partial x}\rho(x,t), \tag{6}$$

where $D$ is given by Eq. (3) and the velocity $v$ is determined by

$$v = \frac{E(\Delta x_i)}{\tau} = (p_r M_r - p_l)\frac{\epsilon}{\tau}. \tag{7}$$

**B. $p_l + p_r \le 1$, $M_r \ge 1$, $M_t \ge 1$, and $M_e \ge 0$ (jump time could be unequal)**

***1. $p_l + p_r = 1$, $M_r \ge 1$, $M_t \ge 1$ (jump time could be unequal and $M_e$ is an unnecessary parameter )***

The asymmetric random walk with $p_l + p_r = 1$ will be derived in this section first; the results will then be generalized to $p_l + p_r \le 1$ in section **II.B.2**. When $p_l + p_r = 1$, there are no immobile sites, and thus the parameter $M_e$ is unnecessary. Because $M_t \ge 1$, the above strategy to calculate variance is difficult to apply to obtain the diffusion coefficient directly. The method employed in Ref. [5], the source point method, will be adopted here to investigate the random walk. Unlike the $M_t = 1$ case in Ref. [5], the total number of jumps in the possible random walk depends on the number of the right jumps that the random walk takes. At time $t = N_t\tau$, if the random walk has total $n_r$ right jumps, then the number of left jumps $n_l$ equals $N_\tau - n_r M_t$, and the total number of jumps is

$$n_l + n_r = N_\tau - n_r(M_t - 1). \tag{8}$$

The probability of finding the particle that has performed $n_l$ left jumps and $n_r$ right jumps is

$$u(n_l, n_r) = (p_l)^{n_l}(p_r)^{n_r}\frac{[N_\tau - n_r(M_t-1)]!}{n_l!\cdot n_r!}. \tag{9}$$

By employing Stirling's approximation [5], Eq. (9) can be approximated as

$$u(n_l, n_r) \approx \sqrt{\frac{N_\tau - n_r(M_t-1)}{2\pi n_l n_r}}\left\{\frac{p_l[N_\tau - n_r(M_t-1)]}{n_l}\right\}^{n_l}\left\{\frac{p_r[N_\tau - n_r(M_t-1)]}{n_r}\right\}^{n_r} = \sqrt{\frac{N_\tau - n_r(M_t-1)}{2\pi n_l n_r}}\,Q(n_l, n_r), \tag{10}$$

where

$$Q(n_l, n_r) = \left\{\frac{p_l[N_\tau - n_r(M_t-1)]}{n_l}\right\}^{n_l}\left\{\frac{p_r[N_\tau - n_r(M_t-1)]}{n_r}\right\}^{n_r}. \tag{11}$$

Applying natural logarithms on $Q(n_l, n_r)$, we have

$$ln[Q(n_l, n_r)] = n_l ln\left\{\frac{p_l[N_\tau - n_r(M_t-1)]}{n_l}\right\} + n_r ln\left\{\frac{p_r[N_\tau - n_r(M_t-1)]}{n_r}\right\}. \tag{12}$$

If the starting site is set as the origin of the coordinate, the position of the particle in the lattice of random walk, $m$, is

$$m = n_r M_r - (N_\tau - n_r M_t) = n_r(M_r + M_t) - N_\tau. \tag{13}$$

From Eq. (13), we have

$$n_r = \frac{m + N_\tau}{M_t + M_r}, \tag{14a}$$

$$n_l = N_\tau - \left(\frac{m + N_\tau}{M_t + M_r}\right)M_t. \tag{14b}$$

By substituting $n_r$ and $n_l$ from Eqs. (14a) and (14b), Eqs. (12) can be rewritten as

$$ln[Q(m,N_\tau)] = \left(\frac{M_rN_\tau - mM_t}{M_t+M_r}\right) ln\left\{\frac{p_l\left[N_\tau - \left(\frac{m+N_\tau}{M_t+M_r}\right)(M_t-1)\right]}{\frac{M_rN_\tau - mM_t}{M_t+M_r}}\right\} + \left(\frac{m+N_\tau}{M_t+M_r}\right) ln\left\{\frac{p_r\left[N_\tau - \left(\frac{m+N_\tau}{M_t+M_r}\right)(M_t-1)\right]}{\left(\frac{m+N_\tau}{M_t+M_r}\right)}\right\} =$$
$$\left(\frac{M_rN_\tau - mM_t}{M_t+M_r}\right) ln(p_l a) - \left(\frac{M_rN_\tau - mM_t}{M_t+M_r}\right) ln(M_rN_\tau - mM_t) + \left(\frac{m+N_\tau}{M_t+M_r}\right) ln(p_r a) - \left(\frac{m+N_\tau}{M_t+M_r}\right) ln(m+N_\tau), \quad (15a)$$

where

$$a = N_\tau(M_r+1) - m(M_t-1). \quad (15b)$$

To find the maximum probability, we perform derivation on $ln[Q(m,N_\tau)]$, which is

$$\frac{\partial}{\partial m}\{ln[Q(m,N_\tau)]\} = \frac{1}{M_t+M_r} ln\left(\frac{M_rN_\tau - mM_t}{p_l a}\right)^{M_t}\left(\frac{p_r a}{m+N_\tau}\right). \quad (16)$$

At $m_{av} = N_\tau\left(\frac{p_rM_r - p_l}{p_rM_t+p_l}\right)$, $a = N_\tau(M_r+1) - m_{av}(M_t-1) = N_\tau\frac{M_t+M_r}{p_rM_t+p_l}$,

$$\{ln[Q(m_{av},N_\tau)]\}' = 0, \quad (17)$$

where the maximum probability occurs. The second derivative of $ln[Q(m,N_\tau)]$ is

$$\frac{\partial^2}{\partial m^2}\{ln[Q(m,N_\tau)]\} = \frac{M_t}{M_t+M_r}\frac{M_t-1}{a} - \frac{M_t}{M_t+M_r}\frac{M_t}{(M_rN_\tau - mM_t)} - \frac{1}{M_t+M_r}\frac{M_t-1}{a} - \frac{1}{M_t+M_r}\frac{1}{(m+N_\tau)}. \quad (18)$$

At $m = m_{av}$, $a = N_\tau(M_r+1) - m_{av}(M_t-1) = N_\tau\frac{M_t+M_r}{p_rM_t+p_l}$, and

$$\frac{\partial^2}{\partial m^2}\{ln[Q(m,N_\tau)]\} = \frac{(M_t-1)^2(p_rM_t+p_l)}{N_\tau(M_t+M_r)^2} - \frac{{M_t}^2(p_rM_t+p_l)}{N_\tau p_l(M_t+M_r)^2} - \frac{(p_rM_t+p_l)}{N_\tau p_r(M_t+M_r)^2} = -\frac{(p_rM_t+p_l)^3}{N_\tau p_l p_r(M_t+M_r)^2}. \quad (19)$$

The third derivative of $ln[Q(m,N_\tau)]$ is

$$\frac{\partial^3}{\partial m^3}\{ln[Q(m,N_\tau)]\} = \left[\frac{M_t}{M_t+M_r}\frac{M_t-1}{a} - \frac{M_t}{M_t+M_r}\frac{M_t}{(M_rN_\tau - mM_t)} - \frac{1}{M_t+M_r}\frac{M_t-1}{a} - \frac{1}{M_t+M_r}\frac{1}{(m+N_\tau)}\right]' = \frac{(M_t-1)^3}{(M_t+M_r)a^2} - \frac{{M_t}^3}{(M_t+M_r)(M_rN_\tau - mM_t)^2} + \frac{1}{(M_t+M_r)(m+N_\tau)^2}, \quad (20)$$

which is negligible compared to the first and second derivatives when $N_\tau$ is large enough. Therefore, by performing the Taylor expansion of $Q(m,N_\tau)$ about $m = m_{av}$ to the second derivative, we obtain

$$ln[Q(m,N_\tau)] \approx -\frac{1}{2!}\frac{(p_rM_t+p_l)^3}{N_\tau p_l p_r(M_t+M_r)^2}(m-m_{av})^2. \quad (21)$$

Note that $\ln[Q(m_{av},N_\tau)] = 0$. From Eqs. (21) and (10), we have

$$u(m,N_\tau) \approx \sqrt{\frac{1}{4\pi\frac{N_\tau p_l p_r}{2(p_rM_t+p_l)}}}\exp\left(-\frac{(m-m_{av})^2}{4\frac{N_\tau p_l p_r(M_t+M_r)^2}{2(p_rM_t+p_l)^3}}\right). \quad (22)$$

By substituting $N_\tau = \frac{t}{\tau}$ and $m = \frac{x}{\epsilon}$, and dividing $\frac{M_t+M_r}{p_rM_t+p_l}$, the probability density is

$$\rho(x,t) \approx \frac{1}{\sqrt{4\pi Dt}}\exp\left[-\frac{(x-vt)^2}{4Dt}\right], \quad (23a)$$

where
$$v = \frac{p_rM_r - p_l}{p_rM_t+p_l}\frac{\epsilon}{\tau}, \quad (23b)$$

$$D = \frac{p_l p_r(M_t+M_r)^2}{2\tau(p_rM_t+p_l)^3}\epsilon^2. \quad (23c)$$

Eqs. (23) satisfies the same type of drift-diffusion equation described by Eq. (6). At $M_t = 1$, Eq. (23c) reduces to $\frac{p_l p_r(1+M_r)^2}{2\tau}\epsilon^2$, which replicated the result in Ref. [5].

2. $\boldsymbol{p_l + p_r < 1}$, $\boldsymbol{M_r \geq 1}$, $\boldsymbol{M_t \geq 1}$, **and** $\boldsymbol{M_e \geq 0}$ ***(jump time could be unequal)***

For $M_t \geq 1$ and $M_e \geq 0$, the average time used for a step in the random walk is

$$\tau_{av} = \gamma\tau, \text{ with } \gamma = p_r M_t + p_l + p_e M_e, \tag{24}$$

where

$$p_e = [1 - (p_l + p_r)] \tag{25}$$

is the probability that the particle stays immobile when starting a new step with the three possible actions: left jump, right jump, or staying still. The average distance that the particle moves in a random walk step is $(p_r M_r - p_l)\epsilon$. Therefore, the velocity of the drift-diffusion is

$$v = \frac{(p_r M_r - p_l)\epsilon}{\tau_{av}} = \frac{p_r M_r - p_l}{p_r M_t + p_l + p_e M_e}\frac{\epsilon}{\tau}, \tag{26}$$

which reduces to Eq. (23b) when $p_l + p_r = 1$ (where $M_e$ is not needed because there are no immobile sites). After obtaining velocity, the remaining task is to get the diffusion coefficient. Let us check the diffusion coefficient in Eq. (23c), which can be rewritten as

$$D = \frac{p_l}{(p_r M_t + p_l)} \cdot \frac{p_r}{(p_r M_t + p_l)} \cdot \frac{(M_t + M_r)^2 \epsilon^2}{2(p_r M_t + p_l)\tau}, \tag{27}$$

From Eq. (3) and the derivation in Section **II.B.1**, we can see that the variance could be determined by how much the random walk deviates from the mean. In Eq. (27), the deviation results from the exchange of the jumps from left to right, or right to left. From Eq. (27), the variance is proportional to

$$\sigma^2 = p_{eff,l}\, p_{eff,r} {\epsilon_e}^2, \tag{28}$$

Where $p_{eff,l} = \frac{p_l}{p_r M_t + p_l}$, $p_{eff,r} = \frac{p_r}{p_r M_t + p_l}$, and $\epsilon_e = M_t \times \epsilon + M_r \epsilon$ is the change in the distance due to the exchange of the jumps. The equation (28) could be approximately generalized to the case where $p_l + p_r < 1$. Similar to Eqs. (2a), for $p_l + p_r < 1$, there are three possible jump exchanges contributing to the total variance: left and right jumps, left jump and non-diffusion, and right jump and non-diffusion. The possible change in diffusion distance, $\epsilon_e$, for exchange between the left and right jumps could be approximated as $M_r\epsilon + M_t \times \epsilon = M_r\epsilon + M_t\epsilon$, or $\epsilon + \frac{1}{M_t} \times M_r\epsilon = \epsilon + \frac{M_r\epsilon}{M_t}$. While for the left jump and non-diffusion, $\epsilon_e$ could be approximated as $0\epsilon + M_e \times \epsilon = M_e\epsilon$, or $\epsilon + \frac{1}{M_e} \times 0\epsilon = \epsilon$ ; and for right jump and non-diffusion, it could be $M_r\epsilon + \frac{M_t}{M_e} \times 0\epsilon = M_r\epsilon$, or $0\epsilon + \frac{M_e}{M_t} \times M_r\epsilon = \frac{M_e}{M_t} M_r\epsilon$. For each type of exchange, one of the two possible $\epsilon_e$ can be selected. In general, the selected $\epsilon_e$ may come from the whole moving length of the long-time action (jump or staying still) plus the scaled moving length of the short-time action (scaled by the time ratios such as $\frac{M_e}{M_t}$). The long-time action can be referred to as the basic action to determine the $\epsilon_e$ for the interexchange. However, in some instances, we will see that the short-time action may be used as the basic action. Based on these three exchanges and $\epsilon_e$, the possible diffusion coefficient could be approximately proposed. Here, a few possible ways to construct the approximate diffusion coefficient are given:

a. $M_e = 1$

When $M_e = 1$, the time for the particle that stays still in the sites is $\tau$. For left and right jumps, the change of particle diffusion distance, $\epsilon_e$ could be approximated as $M_r\epsilon + M_t \times \epsilon = M_r\epsilon + M_t\epsilon$. While for the left jump and non-diffusion, $\epsilon_e$, could be approximated as $1\epsilon + 1 \times 0\epsilon = \epsilon$; and for right jump and non-diffusion, it could be approximated as $M_r\epsilon + M_t \times 0\epsilon = M_r\epsilon$ . Here, for all three exchanges, the long-time actions are selected as the basic actions to contribute to diffusion distance changes due to the exchanges. The total variance from the three exchanges could be approximately given by

$$\sigma^2 \approx \frac{p_l}{\gamma}\frac{p_r}{\gamma}(M_t + M_r)^2\epsilon^2 + \frac{p_r}{\gamma}\frac{p_e}{\gamma}{M_r}^2\epsilon^2 + \frac{p_l}{\gamma}\frac{p_e}{\gamma}\epsilon^2. \tag{29}$$

When $M_t = 1, \gamma = 1$, Eq. (29) reduces to Eq.(3). The diffusion coefficient is

$$D = \frac{\sigma^2}{2\tau_{av}}. \tag{30}$$

b. Small $M_e$

When $M_e$ is not large, the diffusion coefficient could be approximated as

$$D_{smallMe} \approx D_{lr} + D_{le} + D_{re}, \tag{31a}$$

where

$$D_{lr} = \frac{p_l}{\gamma} \cdot \frac{p_r}{\gamma} \cdot \frac{(M_t+M_r)^2 \epsilon^2}{2\gamma\tau} \left(\frac{M_t+M_e}{M_t+1}\right)^3,$$

$$D_{le} = \frac{p_l}{\gamma} \frac{p_e}{\gamma} \frac{(M_e+0)^2 \epsilon^2}{2\gamma\tau} \left(\frac{M_t+M_e}{M_e+1}\right)^3 = \frac{p_l}{\gamma} \frac{p_e}{\gamma} \frac{(M_e\epsilon)^2}{2\gamma\tau} \left(\frac{M_t+M_e}{M_e+1}\right)^3,$$

$$D_{re} = \frac{p_r}{\gamma} \frac{p_e}{\gamma} \frac{\left(\frac{M_t}{M_e} \times 0 + M_r\right)^2 \epsilon^2}{2\gamma\tau} \left(\frac{M_t+M_e}{M_t+M_e}\right)^3 = \frac{p_r}{\gamma} \frac{p_e}{\gamma} \frac{(M_r\epsilon)^2}{2\gamma\tau}. \tag{31b}$$

where $\epsilon_e = M_e \times \epsilon + 0\epsilon$ and $\epsilon_e = M_r \times \epsilon + \frac{M_t}{M_e} \times 0\epsilon$ is selected for the exchanges between left and non-diffusion, and right and non-diffusion, respectively. Here, the long-time action is not used as basic action for obtaining $\epsilon_e$ for the exchange between right jump and non-diffusion if $\frac{M_t}{M_e} < 1$. The selection is based on the simulation results. The diffusion coefficient components in Eq. (31b) are adjusted by $\left(\frac{M_t+M_e}{M_t+1}\right)^3$, $\left(\frac{M_t+M_e}{M_e+1}\right)^3$, and $\left(\frac{M_t+M_e}{M_t+M_e}\right)^3$, which considers that compared to the time $(M_t + M_e)\tau$, the times for the diffusion coefficient components could be approximately scaled by $\frac{M_t+M_e}{M_t+1}$, $\frac{M_t+M_e}{M_e+1}$ and $\frac{M_t+M_e}{M_t+M_e}$.

When $M_t = 1$, and $M_e = 1$, Eq. (31) reduces to Eq. (3). When $M_t > 1$, $M_e \neq 1$, Eq. (31) often gives better results than Eq. (30) based on the simulation results presented in Section IV.

c. Large $M_e$

When $M_e > 3$, the diffusion coefficient could be approximately given by

$$D_{largeMe} \approx D_{lr} + D_{le} + D_{re}, \tag{32a}$$

with

$$D_{lr} = \frac{\frac{p_l}{\gamma_{lr}} \cdot \frac{p_r}{\gamma_{lr}} \cdot (M_t+M_r)^2 \epsilon^2 \left[1 + \frac{\gamma - \gamma_{lr}}{\gamma_{lr}} (p_l + p_r)\right]}{2\gamma\tau},$$

$$D_{le} = \frac{\frac{p_l}{\gamma_{le}} \cdot \frac{p_e}{\gamma_{le}} \cdot (M_e\epsilon)^2 \left[1 + \frac{\gamma - \gamma_{le}}{\gamma_{le}} (p_l + p_e)\right]}{2\gamma\tau},$$

$$D_{re} = \frac{\frac{p_r}{\gamma_{re}} \cdot \frac{p_e}{\gamma_{re}} \cdot \left(\frac{M_e}{M_t} \times M_r \epsilon\right)^2 \left[1 + \frac{\gamma - \gamma_{re}}{\gamma_{re}} (p_r + p_e)\right]}{2\gamma\tau}, \tag{32b}$$

where

$$\gamma_{lr} = p_r M_t + p_l,$$

$$\gamma_{le} = p_l + [1 - (p_l + p_r)]M_e,$$

$$\gamma_{re} = p_r M_t + [1 - (p_l + p_r)]M_e. \quad (32c)$$

In the $D_{re}$ expression, the long-time action is used as the basic action for obtaining $\epsilon_e$ as $\frac{M_e}{M_t} > 1$. In Eq. (32b), the effective probabilities for variance components on the numerators are determined by $\gamma_{lr}$, $\gamma_{le}$, and $\gamma_{re}$, respectively, which could be reasonable considering that at large $M_e$, it may be better to treat the variance for each type of exchange as an independent part of the total variance. The individual variance in Eq. (32b) is adjusted by the corresponding term in the square bracket because the individual average time used for the specific exchange is shorter than the average time for a step in the random walk.

The effect of individual exchange is emphasized in Eq. (32) when $M_e$ is large. Unlike Eq. (31), when $M_t = 1$, and $M_e = 1$, Eq. (32) cannot reduce to Eq. (3).

## III. RANDOM WALK SIMULATION

The discrete random walk simulations were performed to verify the theoretical results obtained in this paper based on the Lattice model [18, 19]. In the simulation, for left and right jumps, their jump lengths are set as $\epsilon$ and $M_r\epsilon$, and their jump waiting periods are $\tau$ and $M_t\tau$, respectively; for the immobile step, the length is set as 0 and the time taken in the step is set as $M_e\tau$. The probabilities for the left and right jumps are $p_l$, and $p_r$, respectively, $p_l + p_r \leq 1$. Although only the displacement at the final recorded time is needed for the simulations, the simulation uses a series of times with a fixed time interval $\tau$ to record the particle position; if the recorded time occurs during a jump step, the recorded position will be adjusted proportionally to the displacement that the unfinished jump step has covered. The simulation counts the frequency of the particle appearing in each range of the spatial coordinate at the last record time of each random walk. The frequency is then divided by the width of the range to give the average frequency. The average frequency divided by the total number of random walks gives the probability distribution function. For each simulation, the time span for each random walk is $10000\tau$, and 20 k repetitions of random walks are used.

## IV. RESULTS AND DISCUSSIONS

The diffusion coefficient is derived for the asymmetric random walk with different parameters: $p_l + p_r \leq 1$, $M_r \geq 1$, $M_t \geq 1$. The results are summarized in Table 1. These results agree well with the discrete random walk simulations. When $p_l + p_r < 1$, the waiting time of the particle in the immobile step $\tau_{im}$ is assumed to be $\tau$. If $\tau_{im} \neq \tau$, $\gamma = p_l + p_r M_t + (1 - p_l - p_r)\frac{\tau_{im}}{\tau}$ will be used in calculating $\tau_{av}$.

Figures 1a-e compare the simulation results with the theoretical predictions for $M_e = 1$. In Figures 1a-b, the asymmetric random walk has $p_l + p_r = 1$, $p_l = 0.25, p_r = 0.75$. In Figure 1a, $M_r = 6$, $M_t = 1$, while in Figure 1b, $M_r = 6$, $M_t = 2$. In Figures 1c-e, the asymmetric random walk with $p_l + p_r = 0.75$, $p_l = 0.25, p_r = 0.5$. In Figure 1c, $M_r = 6$, $M_t = 1$; in Figure 1d, $M_r = 6$, $M_t = 2$; while in Figure 1e, $M_r = 3$, $M_t = 6$. In these Figures, the center of the diffusion peak appears at

$$X_{cen} = vt = \frac{p_r M_r - p_l}{\tau_{av}} t, \quad (33)$$

which is the vertical dash line shown in the center of these figures. From these figures 1a-d, for small $M_t$, $M_t = 1$ or 2, the simulations agree well with their corresponding theoretical predictions. While in Figure 1e, for large $M_t$, $M_t = 6$, the agreement between the simulation and the theoretical prediction is still ok but not as good as that in other Figures 1c-d.

**Table 1**. Drift-diffusion for the discrete asymmetric random walk with different parameters.

| $\frac{\partial}{\partial t}\rho(x,t) = D\frac{\partial^2}{\partial x^2}\rho(x,t) - v\frac{\partial}{\partial x}\rho(x,t), \rho(x,t) \approx \frac{1}{\sqrt{4\pi Dt}}\exp\left[-\frac{(x-vt)^2}{4Dt}\right],\ v = \frac{p_r M_r - p_l}{\tau_{av}}\epsilon,$ $\tau_{av} = \gamma\tau, \gamma = p_r M_t + p_l + p_e M_e$ | | | | |
|---|---|---|---|---|
| $M_e = 1$ *b | $D = \frac{\sigma^2}{2\tau_{av}}, \tau_{av} = \gamma\tau,\ \gamma = p_r M_t + 1 - p_r$ | | | |
| | $M_t \geq 1$ | $M_r \geq 1$ | $p_l + p_r \leq 1$ | $\sigma^2 \approx \frac{p_l}{\gamma}\frac{p_r}{\gamma}(M_t + M_r)^2\epsilon^2 + \frac{p_r}{\gamma}\frac{[1-(p_l+p_r)]}{\gamma}{M_r}^2\epsilon^2 + \frac{p_l}{\gamma}\frac{[1-(p_l+p_r)]}{\gamma}\epsilon^2$ *a |
| | | | $p_l + p_r = 1$ | $\sigma^2 = \frac{p_l}{\gamma}\frac{p_r}{\gamma}(M_t + M_r)^2\epsilon^2$ |
| | $M_t = 1$ | $M_r \geq 1$ | $p_l + p_r \leq 1$ | $\sigma^2 = p_l p_r (M_r + 1)^2\epsilon^2 + p_r[1-(p_l+p_r)]{M_r}^2\epsilon^2 + p_l[1-(p_l+p_r)]\epsilon^2$ |
| | | | $p_l + p_r = 1$ | $\sigma^2 = p_l p_r (1 + M_r)^2\epsilon^2$ [5] |
| | | $M_r = 1$ | $p_l + p_r \leq 1,$ | $\sigma^2 = 4p_l p_r\epsilon^2 + p_l[1-(p_l+p_r)]\epsilon^2 + p_r[1-(p_l+p_r)]\epsilon^2$ [15] |
| | | | $p_l + p_r \leq 1,$ $p_l = p_r$ | $\sigma^2 = (p_l+p_r)\epsilon^2$ [17] |
| $M_e \leq 3$ | $M_t \geq 1, M_r \geq 1$ $p_l + p_r < 1$ | | $D_{smallMe} \approx D_{lr} + D_{le} + D_{re}$, where $D_{lr}$, $D_{le}$, and $D_{re}$ defined by Eq. (31) | |
| $M_e > 3$ | | | $D_{smallMe} \approx D_{lr} + D_{le} + D_{re}$, where $D_{lr}$, $D_{le}$, and $D_{re}$ defined by Eq. (32) | |

a. The expression is generalized based on Eqs. (2a), (3), and (28).

b. When $p_l + p_r = 1$, the parameter $M_e$ is unnecessary.

Figures 2a-e compare the simulation results with the theoretical predictions for $M_e \geq 0$. Both Eqs. (30) and (31) work fine for $M_e = 1$. Eq. (31) gives good results for small $M_e$ in Figures 2a-c. In Figure 2d-e, $M_e > 3$, Eq. (32) best agrees with the random walk simulation. Even in the extreme case, for $M_e = 30$ in Figure 2e, Eq. (32) still gives a reasonable agreement to the simulation. However, in Figures 2d-e, the peaks of the curves predicted from Eq. (30) are much narrower than the simulation peaks, which indicates the corresponding theoretical diffusions are too slow; in contrast, they are too fast from Eq. (31). Therefore, it is important to use an appropriate diffusion coefficient expression to describe a diffusion process.

For $M_t \geq 1$, and $M_e \geq 0$, the diffusion coefficient for the discrete random walk may be obtained based on three possible exchanges: left jump and right jump, left jump and non-diffusion, and right jump and non-diffusion. Each exchange corresponds to a diffusion distance change, $\epsilon_e$, in the random walk. There are two possible $\epsilon_e$ for each type of exchange as explained in Section II.B.2. $\epsilon_e$ may be obtained by the combination of the moving length of the long-time action (jump or staying still) and the moving length of the short-time action scaled by the time ratio. The long-time action should be treated as the basic action for the exchange. However, in Eq. (31), the short time rather than long-time action is used as the basic action for $\epsilon_e$ for the exchange between right jump and non-diffusion when $\frac{M_t}{M_e} < 1$. Additionally, in building the approximate diffusion coefficient expression, some adjustments are necessary because of the time differences existing among these three exchanges. For instance, for small $M_e$ ($M_e \leq 3$ determined from simulation), in Eq. (31), $\left(\frac{M_t+M_e}{M_t+1}\right)^3$, $\left(\frac{M_t+M_e}{M_e+1}\right)^3$, and $\left(\frac{M_t+M_e}{M_t+M_e}\right)^3$ are used to scale the corresponding diffusion coefficient components, while in Eq. (32), for $M_e > 3$, the terms in the square brackets are used to adjust the variance components.

The approximate diffusion coefficient for a discrete random walk with different types of jump sites could be written as

$$D = \sum_{i=1}^{n} \sum_{j=i+1}^{n} D_{ij}, \tag{34}$$

with

$$D_{ij} = c \cdot P_{ieff} \cdot P_{jeff} \cdot \frac{\epsilon_e{}^2}{2\gamma\tau}, \tag{35}$$

where $c$ are the adjusting parameters, $P_{ieff}$ and $P_{jeff}$ are the effective probabilities. In Eqs. (31-32), different c, $P_{ieff}$ and $P_{jeff}$ are used, which depends on the values of $M_t$ and $M_e$. When $M_t$ and $M_e$ are small near 1, the different types of exchanges are more coordinated with each other, and thus Eq. (31) is proposed. In contrast, when $M_t$ and $M_e$ are large, the individual characteristic of each type of exchange becomes strong, and Eq. (32) could provide a better approximation. It still needs more research and simulation efforts to understand how to select the appropriate $\epsilon_e$ and build a better diffusion coefficient expression with the adjusting parameter $c$ and the effective probability parameters: $P_{ieff}$ and $P_{jeff}$.

At each site of the random walk, the particle takes one of three possible actions: left jump, right jump, or staying still. Eqs. (30-32) is obtained based on the exchanges between arbitrary two of these actions. It may be possible that some exchanges involve all three types of actions, when $M_t$ and $M_e$ are large, which needs further research effort to understand how it affects the diffusion coefficient. Further simulations can help us understand the random walk.

When $M_r = 1$, $M_t = 1$, and $M_e = 1$, Ref. [15] has shown that $D = (p_l + p_r)\frac{\epsilon^2}{2\tau} - \frac{v^2}{2}\tau$. When $M_r \neq 1$, $M_t = 1$, and $M_e = 1$, from Eq. (3), $D = \left(p_r M_r{}^2 + p_l\right)\frac{\epsilon^2}{2\tau} - \frac{v^2}{2}\tau$, $v = (p_r M_r - p_l)\frac{\epsilon}{\tau}$, which can be rewritten equivalently as

$$D\tau = \left(p_r M_r{}^2 + p_l\right)\frac{\epsilon^2}{2} - (p_r M_r - p_l)^2\epsilon^2. \tag{36a}$$

The term $(p_r M_r - p_l)^2\epsilon^2$ cannot be neglected regardless of how small $\tau$ value is, which, however, is reduced to zero when $p_r M_r = p_l$. When $M_t \neq 1$ and $M_e = 1$ from Eqs. (29-30), we have

$$D\tau = \left(p_r M_r{}^2 + p_l\right)\frac{\epsilon^2}{2\gamma^2} - \left[p_r^2 M_r{}^2 + p_l^2 - p_l p_r\left(M_t{}^2 + 2M_r M_t - 1\right)\right]\frac{\epsilon^2}{2\gamma^2}. \tag{36b}$$

Discrete random walk and continuous time random walk may have significant differences. From Eq. (23c), when $p_l + p_r = 1$, $M_r = 1$, we have $D = \frac{p_l p_r (M_t+1)^2}{2\tau(p_r M_t + p_l)^3}\epsilon^2$ for discrete random walk. While for CTRW [20], when $p_l + p_r = 1$, $M_r$ =1, the diffusion constant is $D = \frac{\epsilon^2}{2\bar{\tau}}$, $\bar{\tau} = \frac{1}{\frac{1}{2}\left(\frac{1}{\tau_1}+\frac{1}{\tau_2}\right)}$, which is not proportional to $p_l p_r$. In the continuous time random walk (CTRW), the jump time of each step varies, which may result in the fact that near the center of the probability peak of the drift-diffusion, the possibility of the exchange from left to right jumps could be proportional to $p_l$, and the possibility of the exchange from right to left jumps is proportional to $p_r$. Therefore, the diffusion constant does not depend on $p_l p_r$. This significant difference could be employed to determine whether an asymmetric random walk is discrete or continuous, which could be important in real applications. In NMR chemical exchange, the CTRW phase random walk yields a characteristic exchange time twice faster than the discrete random walk [11].

Figure 1.

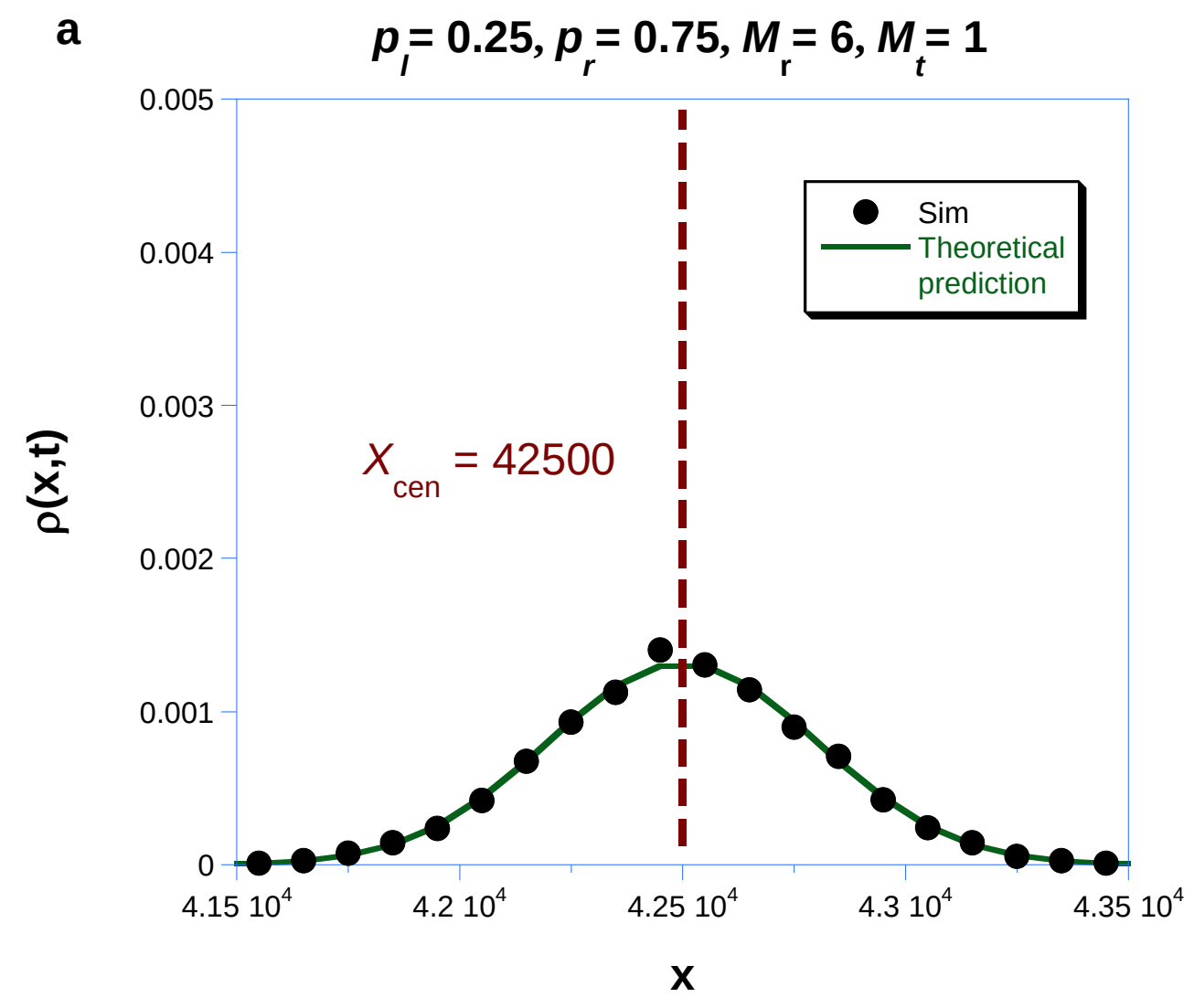

a
$p_l$ = 0.25, $p_r$ = 0.75, $M_r$ = 6, $M_t$ = 1
0.005
0.004
0.003
0.002
0.001
0
ρ(x,t)
Sim
Theoretical prediction
$X_{cen}$ = 42500
4.15 10^4
4.2 10^4
4.25 10^4
4.3 10^4
4.35 10^4
x

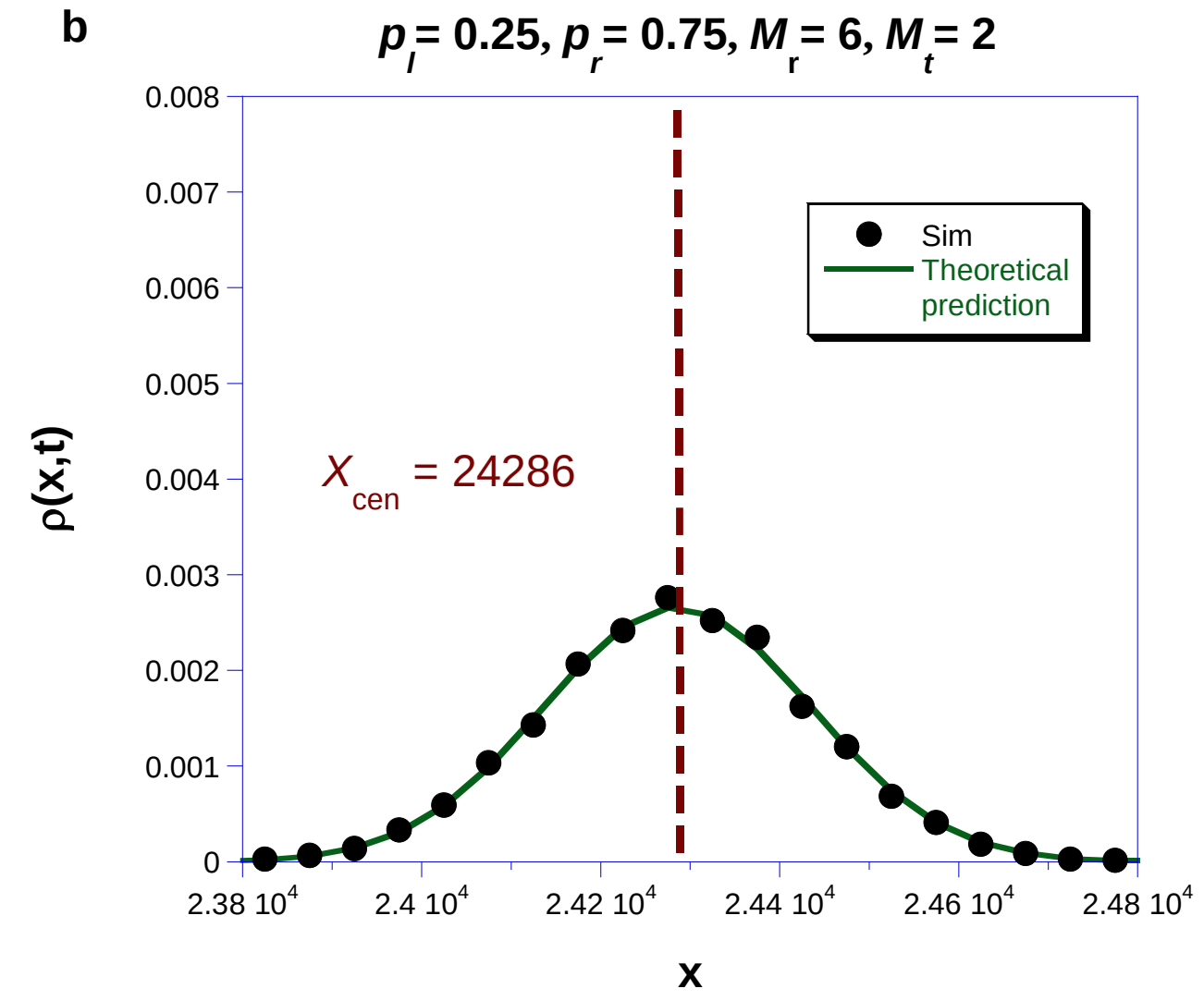

b
$p_l$ = 0.25, $p_r$ = 0.75, $M_r$ = 6, $M_t$ = 2
0.008
0.007
0.006
0.005
0.004
0.003
0.002
0.001
0
ρ(x,t)
Sim
Theoretical prediction
$X_{cen}$ = 24286
2.38 10^4
2.4 10^4
2.42 10^4
2.44 10^4
2.46 10^4
2.48 10^4
x

c

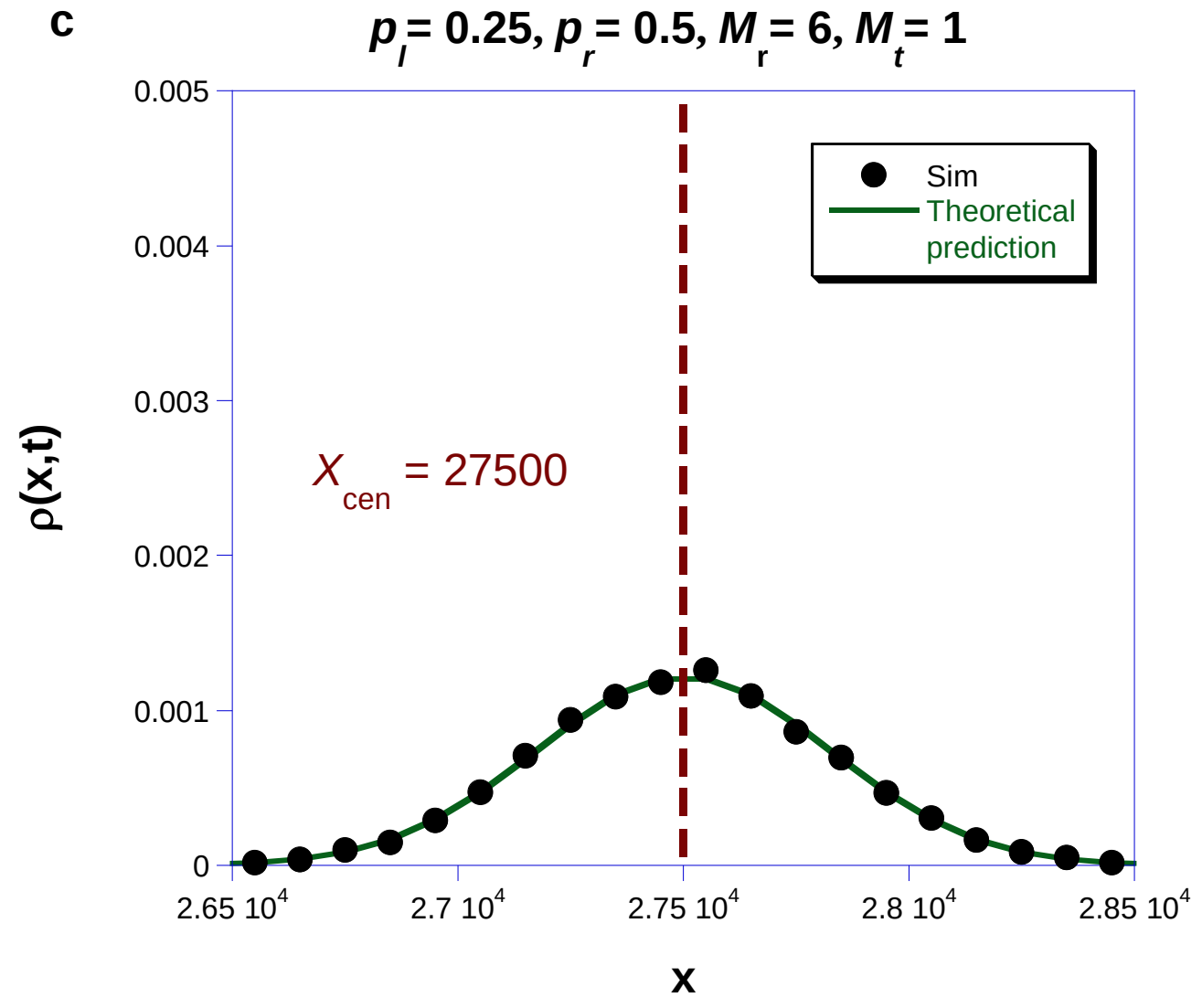
p_l = 0.25, p_r = 0.5, M_r = 6, M_t = 1
0.005
0.004
0.003
0.002
0.001
0
ρ(x,t)
Sim
Theoretical prediction
X_cen = 27500
2.65 10^4
2.7 10^4
2.75 10^4
2.8 10^4
2.85 10^4
x

d

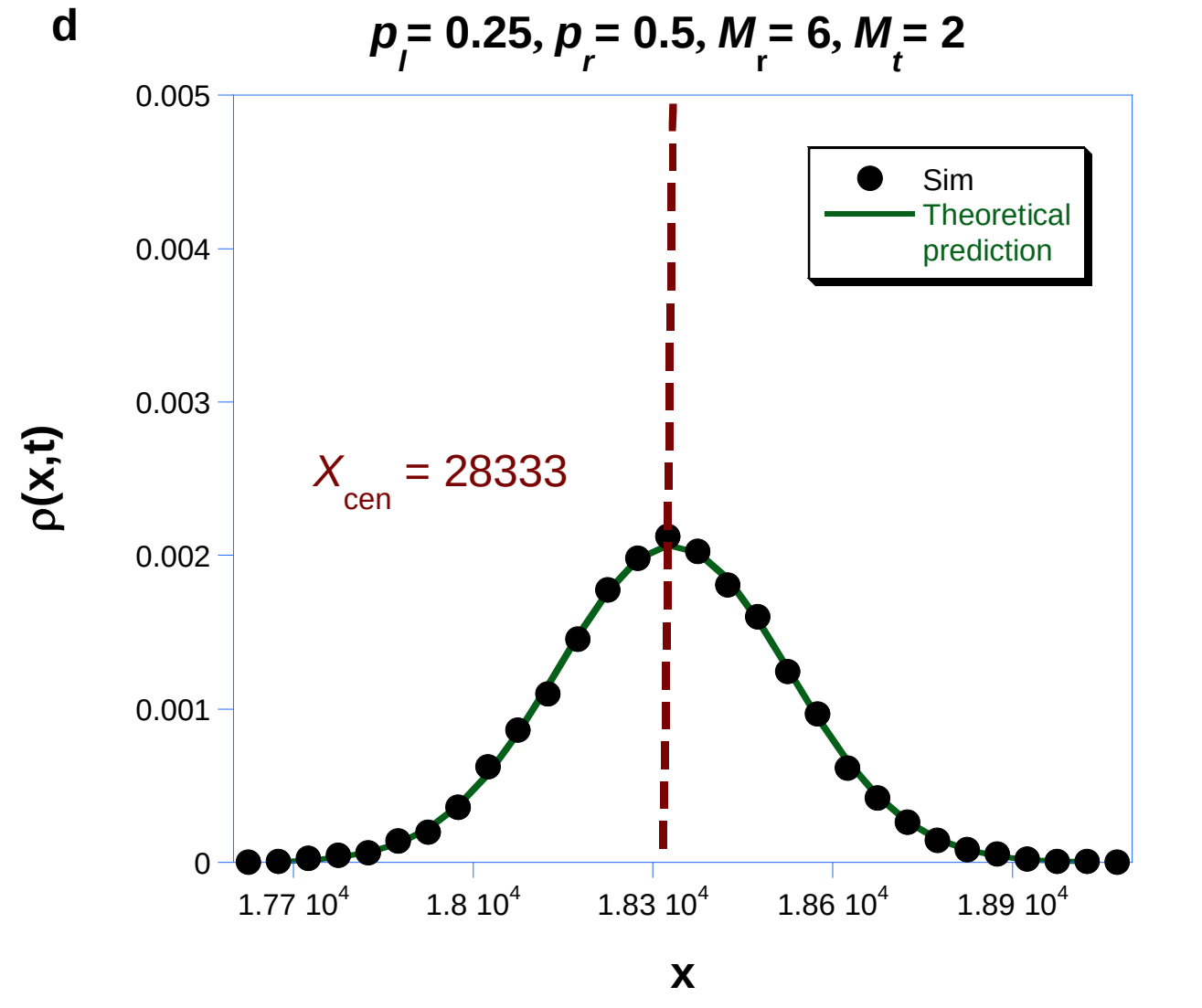
p_l = 0.25, p_r = 0.5, M_r = 6, M_t = 2
0.005
0.004
0.003
0.002
0.001
0
ρ(x,t)
Sim
Theoretical prediction
X_cen = 28333
1.77 10^4
1.8 10^4
1.83 10^4
1.86 10^4
1.89 10^4
x

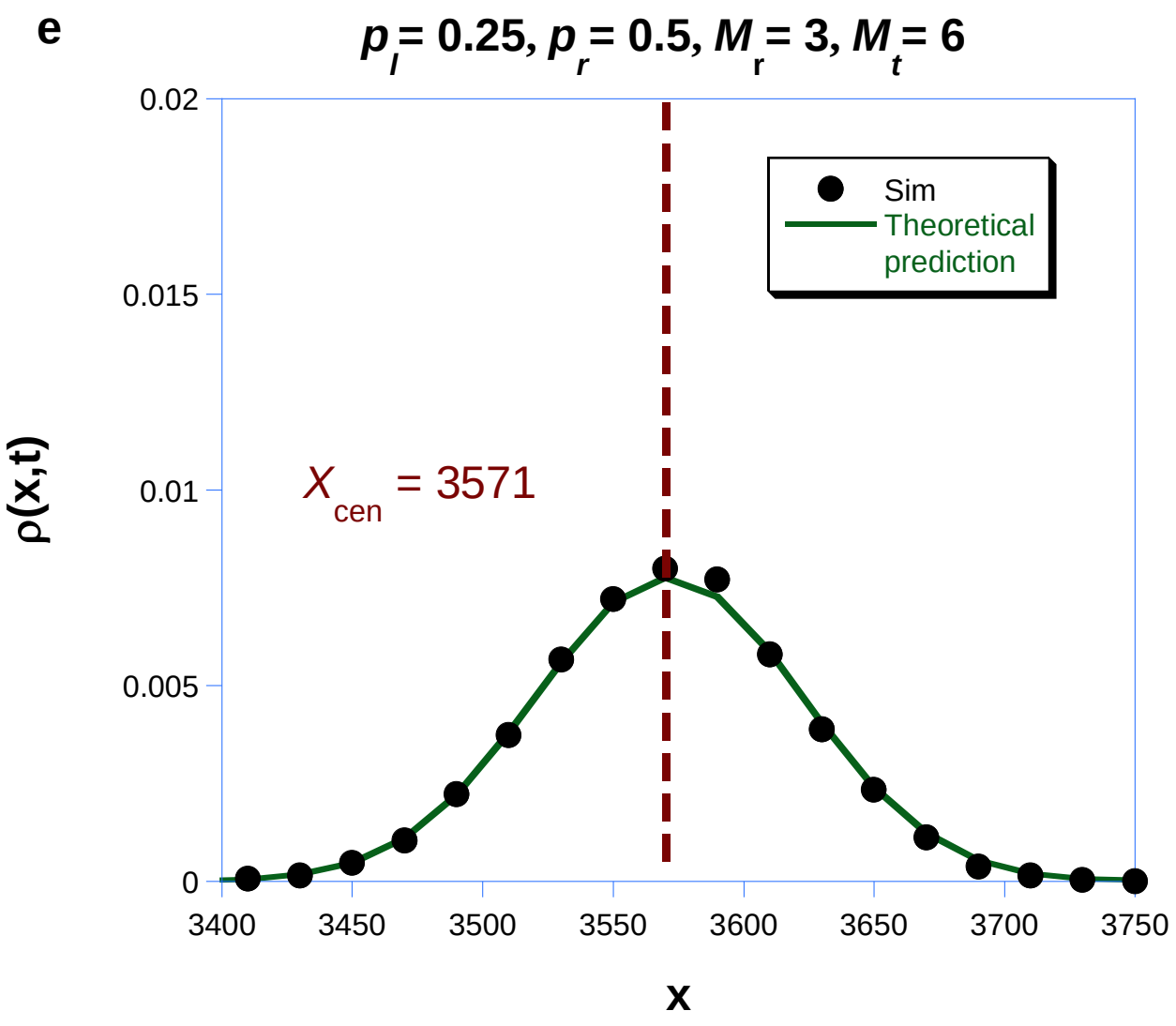

**Figure 1**. Comparison between simulations and theoretical predictions for asymmetric random walk for $M_e = 1$. In 1a-b, $p_l + p_r = 1$, $p_l = 0.25, p_r = 0.75$. In Figure 1a, $M_r = 6$, $M_t = 1$, while in Figure 1b, $M_r = 6$, $M_t = 2$. In Figures 1c-e, $p_l + p_r < 1$, $p_l = 0.25, p_r = 0.5$. In Figure 1c, $M_r = 6$, $M_t = 1$; in Figure 1d, $M_r = 6$, $M_t = 2$; while in Figure 1e, $M_r = 3$, $M_t = 6$. The positions of vertical dash lines occur at $X_{cen} = vt$. The theoretical predictions are drawn based on the expressions listed in Table 1.

Figure 2.

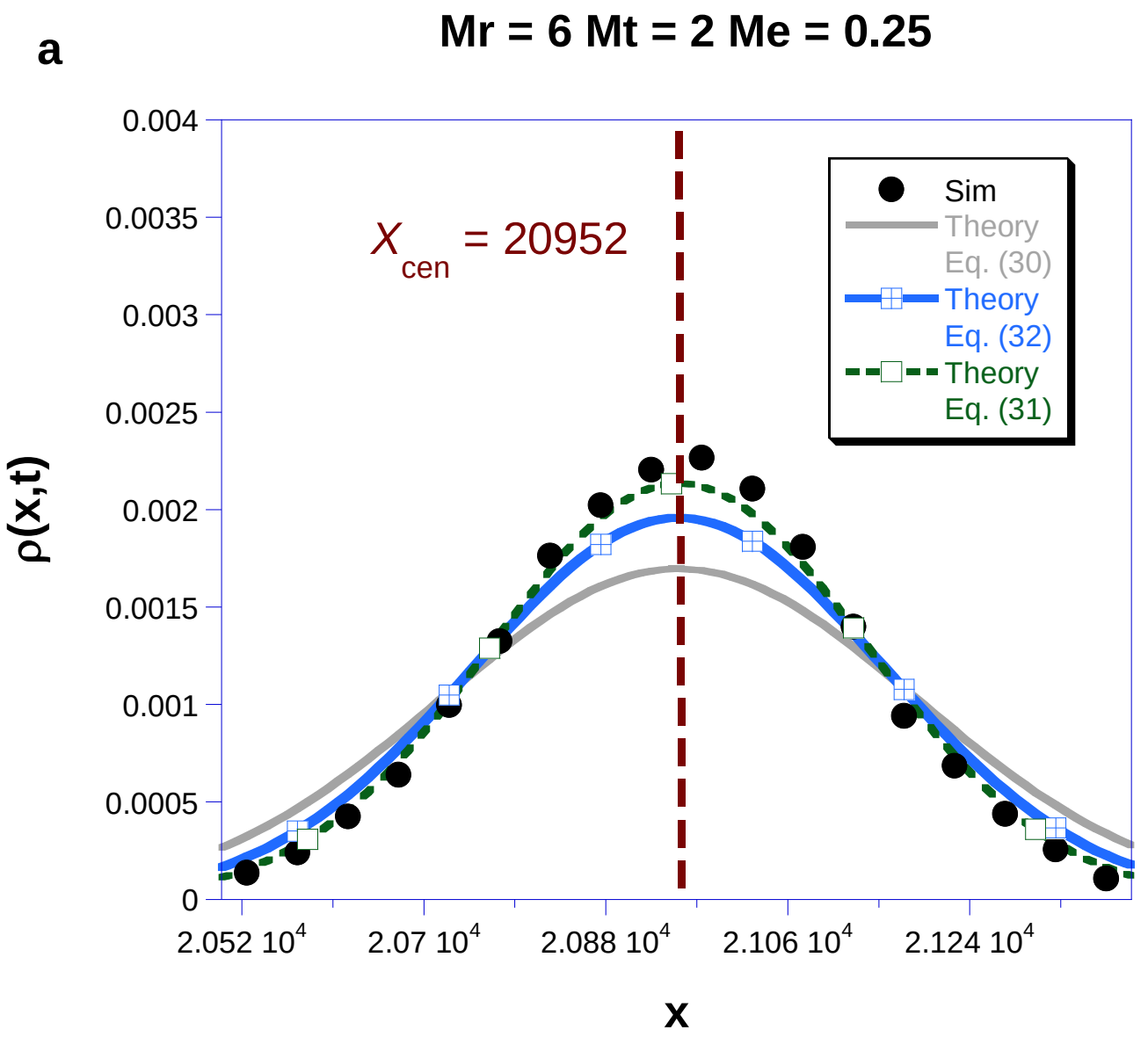

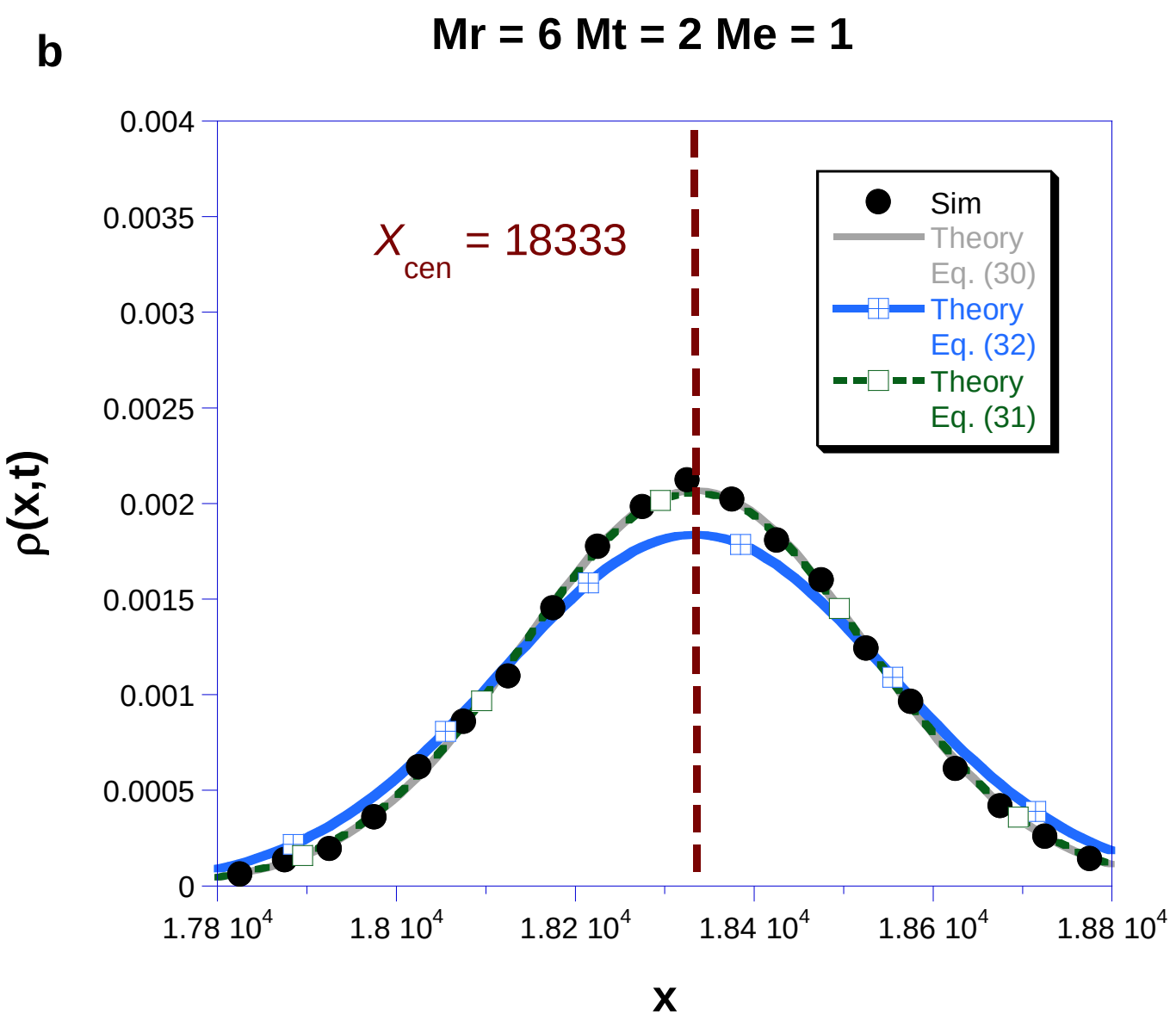
b
Mr = 6 Mt = 2 Me = 1
0.004
0.0035
0.003
0.0025
0.002
0.0015
0.001
0.0005
0
ρ(x,t)
X cen = 18333
Sim
Theory Eq. (30)
Theory Eq. (32)
Theory Eq. (31)
1.78 10^4
1.8 10^4
1.82 10^4
1.84 10^4
1.86 10^4
1.88 10^4
x

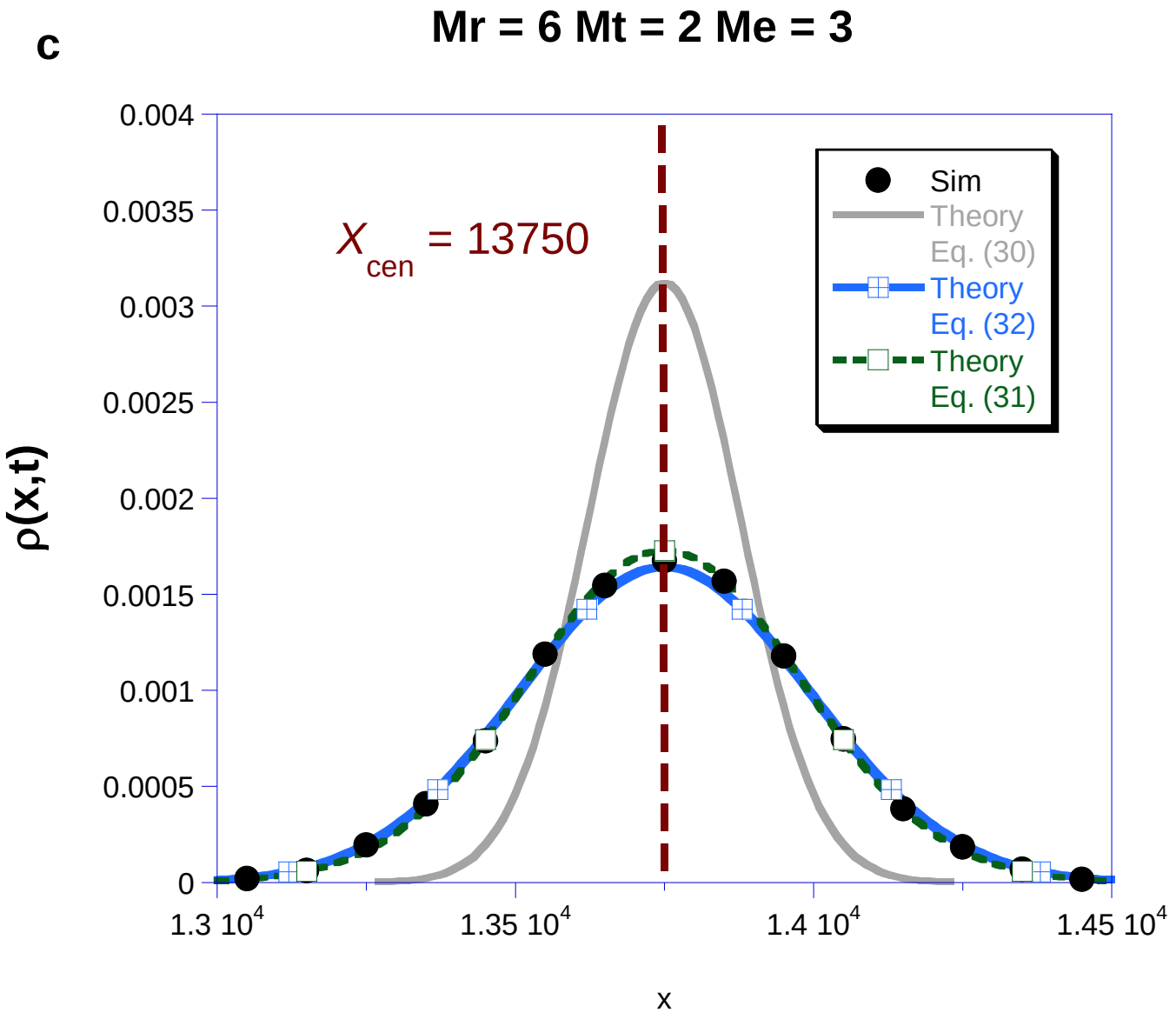
c
Mr = 6 Mt = 2 Me = 3
0.004
0.0035
0.003
0.0025
0.002
0.0015
0.001
0.0005
0
ρ(x,t)
X cen = 13750
Sim
Theory Eq. (30)
Theory Eq. (32)
Theory Eq. (31)
1.3 10^4
1.35 10^4
1.4 10^4
1.45 10^4
x

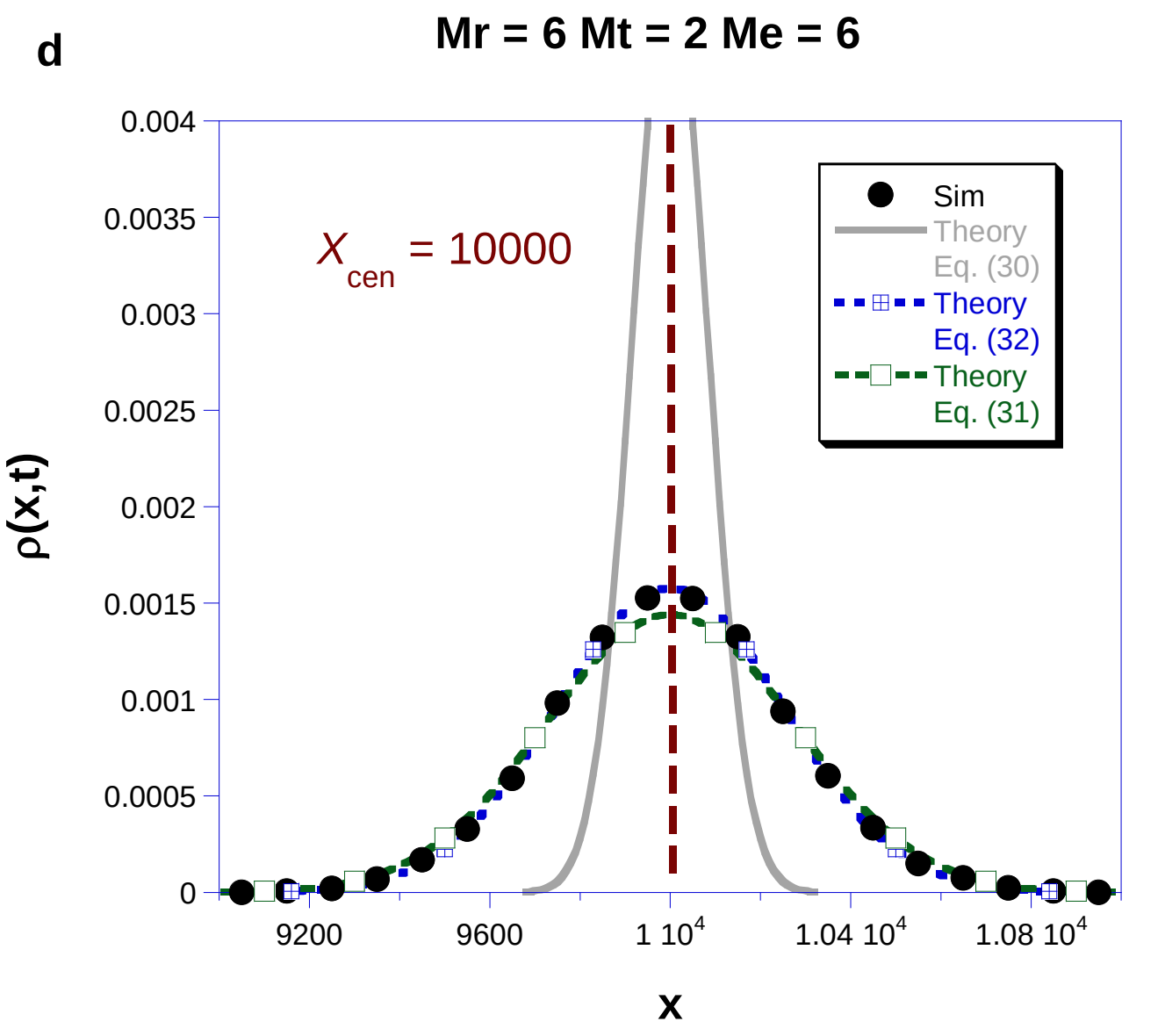

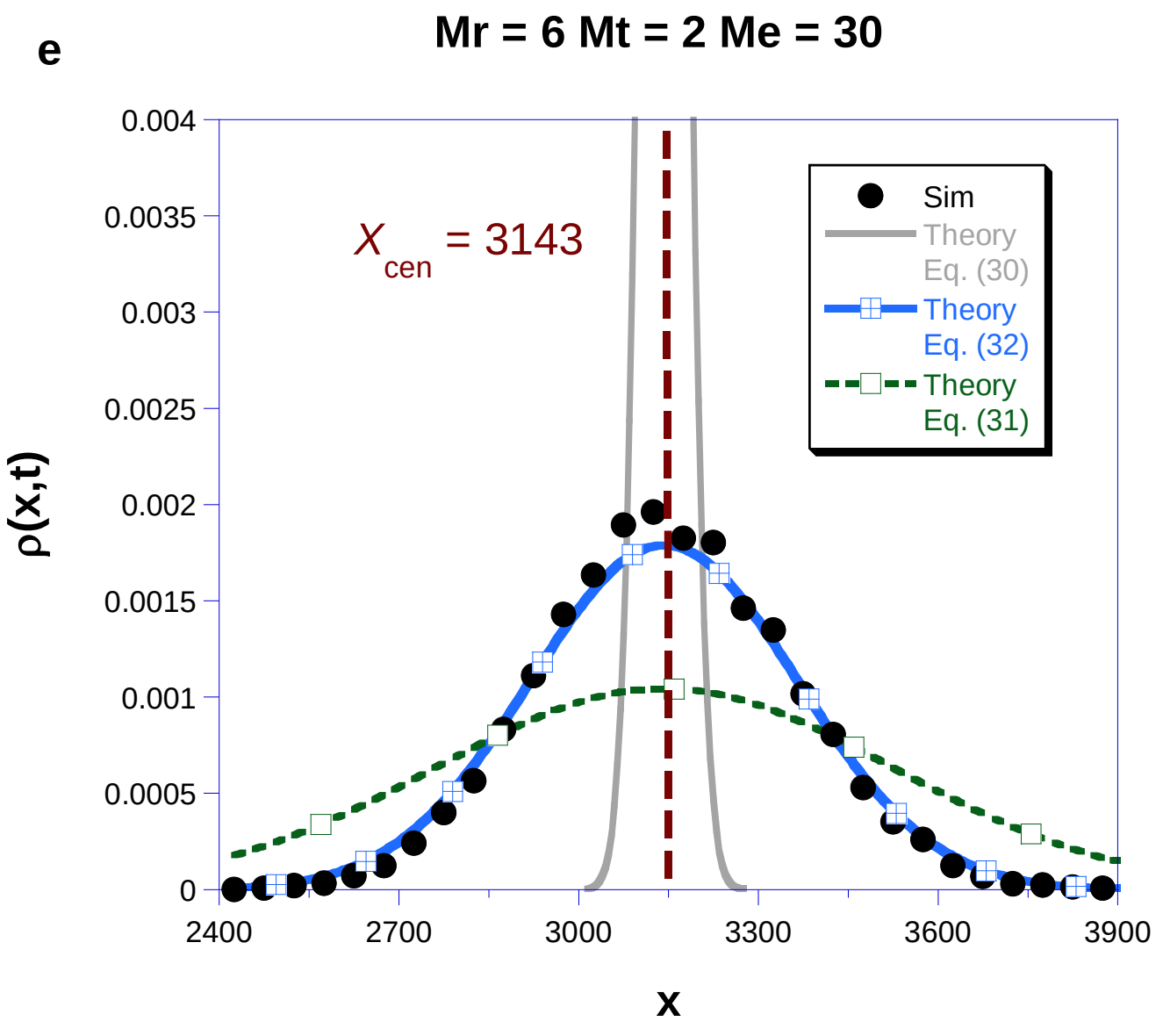

**Figure 2**. Comparison between simulations and theoretical predictions for asymmetric random walk for $M_e \geq 0$ . In all the Figures 2a-e, $p_l + p_r = 1$, $p_l = 0.25, p_r = 0.5$, $M_r = 6$, and $M_t = 2$. The values of $M_e$ are 0.25, 1, 3, 6, 30 in Figures 2a-e, respectively. The positions of vertical dash lines occur at $X_{cen} = vt$. The theoretical predictions are drawn based on Eqs. (30-32).